\documentclass[a4paper,11pt]{article}
\usepackage{jheppub}
\usepackage{graphicx}    
\usepackage{dcolumn}    
\usepackage{bm}           
\usepackage{latexsym}
\usepackage{amssymb}

\title{\boldmath The Blackbody Radiation Laws in the $ \textrm{AdS}_5 \times {\cal S}^5 $ 
Spacetime} 
\author[a]{Ramaton Ramos,}
\affiliation[a]{Centro Brasileiro de Pesquisas F\'{\i}sicas, Rua Dr. Xavier Sigaud 150, 
Urca, RJ 22290-180 -- Brazil}
\author[b]{and Henrique Boschi-Filho}
\affiliation[b]{Instituto de F\'{\i}sica, Universidade Federal do Rio de Janeiro, 
Caixa Postal 68528, RJ 21941-972 -- Brazil}

\emailAdd{ramaton@gmail.com}
\emailAdd{boschi@if.ufrj.br}
\abstract{
In the footsteps of our previous work \cite{RamatonBoschi} we generalize the 
Stefan-Boltzmann and Wien's displacement laws for the $ \textrm{AdS}_5 \times 
{\cal S}^5 $ spacetime, the background of the AdS/CFT correspondence foremost 
realization. 
Our results take into account the $ \textrm{AdS}_5 \times {\cal S}^5 $ full 
dimensionality in the electromagnetic field $A^{\mu}$ wave equation, which 
yields the higher-dimensional blackbody characteristic features suggested in 
literature. In particular, the total radiated power and the spectral radiancy match 
the original Stefan-Boltzmann and Wien's displacement laws in the low-energy 
regime up to available experimental data.}

\begin{document}
\maketitle

\section{Introduction}

The AdS/CFT Correspondence proposed by Maldacena in 1997 relates, in its most 
notable instance, an effective five-dimensional $\textrm{AdS}_5$ gravitational 
theory (via type IIB string theory) to a four-dimensional Conformal Field Theory 
($N=4$ supersymmetric Yang-Mills theory) on its boundary \cite{Maldacena98}, 
being the subject of interesting contemporary physical unfoldings 
\cite{Itzhaki:1998dd,Maldacena:1998im,Gubser1998PLB,WittenATMP98,
KlebanovWitten,Aharony:1999ti,Hawking:2000da,semiclassicalduality02,
RyuTakayanagi06,pomeronPolchinskiStrassler,gravityfluids,Lin:2004nb,
Berenstein:2005aa,Gaiotto:2009gz}. 

The duality expressed in this correspondence is the most successful realization of 
the Holographic Principle \cite{Susskind1995,WittenholoAdS98,Susskind:1998dq,
Bousso2002}, which enables to ``translate'' certain problems in nuclear and 
condensed matter physics into more tractable ones in the string theory picture
\cite{Herzog:2009xv,Alday:2007hr,Herzog:2006gh,SonStephanov05,SonLett01}.
Albeit the theories involved are not viable models of the real world, they provide 
model-building tools for physics beyond the standard model and some of their 
properties elucidate important questions, such as the quantum behavior of black 
holes \cite{Horowitz:1996ay,Maldacena:1996gb,Horowitz:1996ac,Maldacena:1997ih,
Maldacena:1998bw,Maldacena:2001kr,Horowitz:2003he,Maldacena2014ova}.

Anti-de Sitter space is the maximally symmetric Einstein equations solution with 
negative (attractive) cosmological constant. In our universe, this constant is certainly 
not attractive, yet it can be regard as a long-distance regularization of gravity behavior. 
$ \textrm{AdS}_5 $ space notable feature relates to its boundary, which looks locally 
around each of its points like our well-known Minkowski ${\cal M}^{1,3}$ spacetime, 
labeled from now on as ${\cal M}$ unless otherwise specified. The five additional 
``compact'' dimensions, ${\cal S}^5$ subspace, come as a dimensional requirement for 
String Theory mathematical consistency.

Blackbody radiation is described by Planck's law, which implies its radiative features. 
In particular, the Stefan-Boltzmann law predicts that the blackbody radiated power is 
$ R(T) = \sigma_{_B} T^4 $, with $\sigma_{_B}$ as the Stefan-Boltzmann constant, 
while the Wien's displacement law relates the blackbody temperature $T$ to the 
wavelength $\lambda_m$ corresponding to the wavelength radiancy $R(T,\lambda)$ 
peak, $\lambda_m\,T = 2.897 \cdot 10^{-3} \, \rm m\,K$. 

These laws shall depend on the spacetime dimensionality 
\cite{Alnes:2005ed,Cardoso:2005cd}, in contrast with observed data. Our results 
preserve the well-known blackbody radiation laws in the low-temperature regime 
while obtaining the higher-dimensional theoretical relations in the high-temperature 
regime, explicitly establishing a connection between the current theoretical 
unfoldings and our well-acquainted $4$-dimensional experience.

\vspace{.1cm}
The article is organized as follows. In section 2 we lay the blackbody concept and 
briefly introduce our approach. In Section 3 we single the Minkowski slice out of the 
$\textrm{AdS}_5$ space and generalize the Stefan-Boltzmann and Wien's laws 
for the specific $ {\cal M} \times {\cal S}^5 $ scenario. Similar steps are taken in 
section 4 towards the full $ \textrm{AdS}_5 \times {\cal S}^5 $ spacetime 
generalizations. In section 5 we set experimental bounds on the $\textrm{AdS}_5$ 
conformal coordinate, plotting our main results for these bounds. 
The final section is devoted to closing remarks and general comments.

\section{The Blackbody and our Approach}

A blackbody is defined as a body with a rich energy spectrum, capable of exciting 
all frequencies of light by thermalization. As consequence, all blackbodies at the same 
temperature emit thermal radiation with the same spectrum. For a current review on 
the matter, see \cite{Lehoucq2011}. 

The standard textbook blackbody approach consists in taking a small bidimensional 
orifice connecting an isothermal enclosure to its outside as a blackbody surface, a 
more technical and conceptual discussion on approximating a blackbody for a 
blackbox is found in \cite{Garcia-Garcia2008,Smerlak2011}. 

Here, we look for the electromagnetic wave equation solution which fills 
the $ \textrm{AdS}_5 \times {\cal S}^5 $ spacetime, where boundary 
conditions are assumed, in order to obtain the radiated spectrum. The general 
procedure employed in this work is analogous to that of \cite{RamatonBoschi}.

The $ \textrm{AdS}_5 \times {\cal S}^5 $ spacetime metric can be written in the form
\begin{equation}\label{metricAdS5xS5}
{ds}^2 = R^2 \left[ z^{-2} \left( {dx}^{\mu} {dx}_{\mu} + {dz}^2 \right) 
+ {d \Omega}_5^2 \right] \, ,
\end{equation}
and the wave equation to be satisfied by the electromagnetic field $A^{\mu}$ 
for a generic metric $g_{ab}$ is  
\begin{equation}\label{generalwaveequation}
{(-g)}^{-\frac{1}{2}} {\partial}_{a} \left[ {(-g)}^{\frac{1}{2}} g^{ab} 
{\partial}_{b} A^{\mu} \right] = 0 \, ,
\end{equation}
where the $a$ and $b$ indices run through all considered dimensions while we 
keep the $\mu$ indices for the Lorentzian polarizable ones.

Making use of Bose-Einstein statistical prescription and accounting two helicity states 
related to the propagative aspect of the photons associated to the electromagnetic 
waves, the blackbody radiation energy density at temperature $T$ is 
\begin{equation}\label{densityenergy}
{\rho}(T) = \frac{2}{V} \sum_i \frac{h{\nu}_i}{e^{h{\nu}_i/kT}-1} \ ,
\end{equation}
where $V$ is taken as the ordinary $3$-dimensional volume, since this is our usual 
framework and the $ \textrm{AdS}_5 \times {\cal S}^5 $ metric is scale invariant.

The spatial boundary conditions determine the photons energy states $h {\nu}_i$, 
and the energy density is proportional to the radiancy $R(T)$, the energy rate per 
unit area, through a geometric speed-of-light factor, 
\begin{equation}\label{RTotal}
R (T) = \frac{c}{4} \, \rho (T) \ .
\end{equation}

\section{$ {\cal M} \times {\cal S}^5 $ brane scenario}

The present $ {\cal M} \times {\cal S}^5 $ setup, in which the conformal 
coordinate $z=\textrm{const}$, has effective metric
\begin{equation}\label{metricMxS5}
{ds'}^2 = {dx}^{\mu} {dx}_{\mu} + z^2 {d \Omega}_5^2 \, ,
\end{equation}
which means that the electromagnetic field fulfills 
\begin{equation}\label{MxS5waveeq}
\left[ {\partial}^{\nu} {\partial}_{\nu} + \frac{1}{z^2} {\Delta}_{{\cal S}^5} 
\right] A^{\mu} (x,\Theta) = 0 \, ,
\end{equation}
where ${\Delta}_{{\cal S}^n}$ is the Laplace-Beltrami operator on ${\cal S}^n$ 
with the particular property 
\begin{equation}\label{sphericharmonic}
\left[\Delta_{{\cal S}^n} + l(l+n-1) \right] \, Y_l ({\Theta}) = 0 
\end{equation}
and $  Y_l ({\Theta}) = Y_{l_n \cdots l_1} ({\theta}_n,\cdots, {\theta}_1) $ are 
the spherical harmonics in higher dimensions, with $ l \doteq l_n $. Since $  l_5 
\geq l_4 \geq l_3 \geq l_2 \geq |l_1| \geq 0 $, for each $l=0,1,2,...$, one has 
\begin{equation}\label{degenerL}
d_l = \frac{1}{4!}(l+1)(l+2)(l+3)(2l+4)
\end{equation}
associated degeneracies.

\vspace{0.1cm}
Solution of (\ref{MxS5waveeq}) is given by variable separation, $ \textrm{\bf 
A}^{\mu} $ indicates the field polarization, thus
\begin{equation}\label{MinkowskiAfield}
A^{\mu} (x,\Theta) = \textrm{\bf A}^{\mu} e^{i k \cdot x} \, Y_l ({\Theta}) \, ,
\end{equation} 
from which we obtain, by Dirichlet boundary conditions, 
\begin{equation}\label{MxS5specter}
{\left( \frac{2\pi\nu}{c} \right)}^2 = \displaystyle \sum_{i=1}^{3} n_i^2 
{\left(\frac{\pi}{L_i}\right)}^2 + \frac{l(l+4)}{z^2} \, .
\end{equation}
where $L_i$ are the usual three-dimensional coordinate lengths and $n_i=0,1,2,...$ 

Since the ${\cal M}$ spatial and ${\cal S}^5$ contributions may not be on the 
same foot, particularly for $z \ll L_i$, eq.(\ref{RTotal}) is better written via 
(\ref{densityenergy}) as 
\begin{equation}\label{lseparation}
R (T) = \frac{c}{2V} \left[ \sum_{l=0} + \sum_{l \neq 0}
            \right] \frac{h \nu_{il}}{e^{h\nu_{il}/{kT}}-1} \ .
\end{equation}

\vspace{1mm}
The first sum is a single one in $n_i$. Taking it by an integral and considering 
$\nu_{_0} \equiv \nu(n_i,0)$ yields
\begin{equation}\label{SBdefl0}
R_{0} (T) \doteq \frac{c}{2V} \sum_{l=0} \frac{h\nu_{_0}}
           {e^{h\nu_{_0}/{kT}}-1} = \int_{0}^{\infty} R_0(T,\nu_{_0})\,d\nu_{_0}
\end{equation}
where the spectral radiancy $R_0 (T,\nu_{_0})$ for this range of modes is 
\begin{equation}\label{R0nu}
R_0 (T,\nu_{_0}) = \frac{2 \pi h}{c^2} 
                            \frac{{\nu_{_0}}^3}{e^{h \nu_{_0}/kT}-1} \,.
\end{equation}
 
Making the variable change $ y_{_0} = h\nu_{_0} / kT $ one integrates 
(\ref{R0nu}) arriving at
\begin{equation}
R_0 (T) = \frac{2\pi c}{{(hc)}^3} {(kT)}^4 \int_{0}^{\infty} 
\frac{y_{_0}^3}{e^{y_{_0}}-1} \, dy_{_0} \ ,
\end{equation}
with the integral expressed by the mathematical identity 
\begin{equation}\label{zetagama}
\int_{0}^{\infty}\frac{x^d}{e^x-1}\,dx = \Gamma (d+1) 
\, \zeta (d+1) \ .
\end{equation}

Taking $ d=3 $ in the above identity, one obtains the ordinary radiancy contribution 
\begin{equation}\label{R0}
R_0 (T) = \sigma_{_B} \, T^4 \ ,
\end{equation}
which is the well-known Stephan-Boltzmann law with
\begin{equation} 
\sigma_{_B} = \frac{2{\pi}^5 k c}{15} {\left( \frac{k}{hc} \right)}^3 = 
\, 5.67 \cdot 10^{-8} \, {\rm W} \, {\rm m}^{-2} {\rm K}^{-4} \, .
\end{equation}

\vspace{1mm}
Now we consider the case $ l \neq 0 $ in which both $n_i$ and $l$ contribute to 
the spectral radiancy. Taking the sum with $\nu \equiv \nu(n_i,l)$, 
one gets
\begin{equation}\label{Rldef}
R_{l} (T) \doteq \frac{c}{2V} \sum_{l \neq 0} d_l \, 
           \frac{h\nu}{e^{h\nu/{kT}}-1} \ .
\end{equation}

Note that the $d_l$ degeneracies given by eq.(\ref{degenerL}) are related to 
the frequencies through eq.(\ref{MxS5specter}). Considering the mean values 
of this dispersion relation we get
\begin{equation}
\langle \frac{l(l+4)}{z^2} \rangle = \frac{1}{4} {\left( \frac{2\pi\nu}{c} \right)}^2 \, ,
\end{equation}
which for high $l$ values allows the substitution $d_l \rightarrow d_l(\nu)$ with 
$l = {\pi z \nu}/{c}$. Then, taking the sum by an integral and considering just 
the $d_l(\nu)$ leading order contribution one gets
\begin{equation}\label{Rlint}
R_{l} (T)  = \int_{0}^{\infty} R_{l}(T,\nu) \, d\nu \ ,
\end{equation}
where the corresponding spectral radiancy is given by
\begin{equation}\label{R9nu}
R_{l}(T,\nu) = \frac{2}{4!} {\left(\frac{\pi z \nu}{c} \right)}^4 \,
                      \frac{{\Omega}_{(4)} \, \pi z \, h c}{2 \, c^4}  
                     \frac{{\nu}^{4}}{e^{h\nu/kT}-1} \ \, ,
\end{equation}
and ${\Omega}_{(d)} = 2 \, {\pi}^{d/2} / \, \Gamma (d/2)$ is the solid angle 
in a $d$-dimensional space. 

Through the variable change $ y = h\nu / kT $ one gets
\begin{equation}
R_{l} (T) = \frac{{\Omega}_{(4)}}{3} \frac{hc \,{(\pi z)}^5}{8 \, c^8}  
                {\left( \frac{kT}{h} \right)}^9 \int_0^{\infty} \frac{y^8}{e^y-1} \, dy \ ,
\end{equation}
and using (\ref{zetagama}) the radiancy contribution $R_{l}(T)$ due to these modes is 
\begin{equation}\label{R9}
R_{l} (T) = \sigma_{l} \, T^9 \ ,
\end{equation}
\begin{equation}\label{sigma9}
\sigma_{l} \, = \frac{{\Omega}_{(4)} k c }{24} \, 
         {\left( \frac{k}{hc} \right)}^{8} 
         \Gamma(9) \, \zeta(9) \ {(\pi z)}^5 \ .
\end{equation}

\vspace{.1cm}
Grouping the computed radiancy contributions one gets for the total blackbody energy 
rate per unit area 
\begin{equation}\label{RTotal9}
R(T) = \sigma_{_B} \, T^4 + \sigma_{l} \, T^9 \ ,
\end{equation}
which is interpreted as the generalized Stefan-Boltzmann law for the $ {\cal M} 
\times {\cal S}^5 $ subspace.

The nature and validity of the preceding outcome is closely related to the 
temperature of the blackbody in question. For low 
temperatures the obtained Stefan-Boltzmann law generalization (\ref{RTotal9}) 
reduces to its well-known form (\ref{R0}).

On the other side, for sufficiently high temperatures, $ T \gg {hc}/{kz} $, all 
considered dimensions match up on equal footing from the perspective of the 
actual irradiative process. Then, the generalized Stefan-Boltzmann law takes on its 
higher-dimensional character $ R(T)=\sigma_{l} \, T^9 $, which agrees 
with the results in \cite{Alnes:2005ed,Cardoso:2005cd}. 

\vspace{.2cm}
To obtain the respective Wien's law we write the total wavelength radiancy as 
\begin{equation}\label{Rlambda}
R(T,\lambda) = R_0(T,\lambda)+R_{l}(T,\lambda) \ ,
\end{equation}
where 
\begin{eqnarray}
R_0(T,\lambda) &=& \frac{2 \pi h c^2}{\lambda^5} \, 
                     {(e^{\frac{hc}{k T \lambda}}-1)}^{-1} \ ,
\\
R_{l}(T,\lambda) &=& \frac{\Omega_{(4)} h c^2}{24 \, \lambda^{10}} \, 
                     {(\pi z)}^5 {(e^{\frac{hc}{k T \lambda}}-1)}^{-1} \ .
\end{eqnarray}

Since for a given $T$ there is a $\lambda_{_m}$ for which $R(T,\lambda_{_m})$ 
is maximum, through $dR(T,\lambda)/d\lambda = \, 0$ one obtains 
\begin{equation}\label{gwien9}
1 - e^{-{hc/kT\lambda_{_m}}} = \frac{hc}{k T \lambda_{_m}} \, 
                              \frac{1+\epsilon_{l}(z,\lambda_{_m})}
                              {5+10\,\epsilon_{l}(z,\lambda_{_m})} \ ,
\end{equation}
where $\epsilon_{l} (z,\lambda) = {R_{l}(T,\lambda)}/{R_0(T,\lambda)}$ 
is defined as the corresponding {\it wavelength radiancy relative deviation}, 
\begin{equation}\label{wavelengthdeviation9}
\epsilon_{l} (z,\lambda)  = \frac{4 \pi}{3}
                                      {\left(\frac{\pi z}{2 \lambda}\right)}^{5} \ .
\end{equation}

Eq.(\ref{gwien9}) is regarded as the generalized Wien's law for $ {\cal M} 
\times {\cal S}^5 $ subspace, once it relates the blackbody temperature $T$ with 
the wavelength $\lambda_m$ corresponding to the maximum value of the total 
wavelength radiancy $ R(T,\lambda)$.

For wavelengths much larger than the conformal coordinate value, 
$ \lambda_m \gg z $, Eq. (\ref{gwien9}) reduces to $ 1 - e^{-x_4} =  x_4 / 5 $
($x_4 \equiv {h c} / {k T \lambda_m} $), which implies the usual Wien's 
displacement law, namely $ \lambda_m \, T = 2.897 \cdot 10^{-3} \, \rm m\,K$. 

On the opposite limit, for $\lambda_m \ll z $, Eq.(\ref{gwien9}) reduces to 
\begin{equation}\label{lowgwien}
 1 - e^{-x_{9}} = \frac{1}{10} \, x_{9} \quad \;; 
\quad x_{9} \equiv {\frac{h c}{k T \lambda_m}}
\end{equation}
where $x_{_{D}}=W(-(D+1)e^{-(D+1)})+D+1$, and $W(y)$ is the Lambert 
function. This yields $ \lambda_m \, T = x_{9} \, hc/k$ and the generalized Wien's 
displacement law assumes its higher-dimensional behavior.

\section{The $ \textrm{AdS}_5 \times {\cal S}^5 $ scenario}

For the present case in which our spacetime is the full $ \textrm{AdS}_5 \times 
{\cal S}^5 $, the considered electromagnetic fields must satisfy the equation
\begin{equation}\label{AdSwaveeq}
\left[ {\partial}_{z}^2 - \frac{3}{z} {\partial}_z + {\partial}^{\nu} {\partial}_{\nu}
+ \frac{1}{z^2} {\Delta}_{{\cal S}^5} \right] A^{\mu} (x,\theta_i,z) = 0 \, ,
\end{equation}
whose solution is 
\begin{equation}\label{AdSfield}
A^{\mu} (x,\Theta,z) = \textrm{\bf A}^{\mu} \, e^{i k \cdot x} \, 
Y_l({\Theta}) \, Z(z) \, ,
\end{equation} 
such that the conformal dependence is
\begin{equation}\label{Zsol}
Z (z) = z^2 \left( c_1 \, J_{l+2} (\kappa z) + c_2 \, N_{l+2} (\kappa z) \right) \, ,
\end{equation}
where $ J_m (x) $ and $ N_m (x) $ are respectively the Bessel and Neumann 
functions of order $m$. 

\vspace{.1cm}
The above pair of functions emerge as radial solutions to the Helmholtz and Laplace's 
equations in cylindrical coordinates, thus sharing similar properties. They are quite 
important for many problems, like the ones involving electromagnetic waves in 
cylindrical waveguides and frequency-dependent friction in circular pipelines.

For integer $m$ and $x \in \mathbb{R}$, $J_m(x)$ is finite at the origin 
$x=0$ while $N_m(x)$ is not. Several properties of the Bessel (and Neumann) 
function are derivable from the following remarkable definition, 
\begin{equation}
J_m(x) = \frac{1}{\pi} \int_{0}^{\pi} \cos{(x\sin{\tau}-m\tau)}\,d\tau \, .
\end{equation}

Alternatively, one can work with its Taylor series expansion, $\alpha \in \mathbb{R}$,
\begin{equation}
J_{\alpha} (x) = \sum_{k=0}^{\infty} \frac{(-1)^k}{k! \, \Gamma(k+\alpha+1)} 
{\left(\frac{x}{2}\right)}^{2k+\alpha} \, .
\end{equation}
While the Neumann function is suitably described by 
\begin{equation}
N_{\alpha} (x) = \frac{J_{\alpha}(x) \cos(\alpha \pi) - J_{-\alpha} (x)}
{\sin(\alpha \pi)} \, ,
\end{equation}
valid for integer $\alpha$ as a limiting case.

Our motivation behind the introduction of boundary conditions 
in the present setup is the Randall-Sundrum model \cite{Randall:1999} 
(with its subsequent modifications such as \cite{Goldberger:1999wh},
where bulk fields are considered) for which the space is an orbifold 
consisting of two glued portions of the $ \textrm{AdS}_5 $. To assure 
continuity of fields in its bulk one is led  to Neumann boundary conditions 
at its two interfaces. 

Our choice for the Dirichlet boundary conditions at these interfaces is feasible 
since we are going to sum over an infinite tower of modes and the difference between
these two boundary conditions affect only the shape of the considered modes, but not 
their corresponding wavelengths. 

Setting $a$ and $b$ as the boundary conformal coordinates for which (\ref{Zsol}) 
matches Dirichlet conditions, $Z(a)=Z(b)=0$, determines the possible $\kappa$ 
values that will fit into the spectral relation derived from (\ref{AdSwaveeq}) and 
(\ref{AdSfield}),
\begin{equation}\label{spec10}
\frac{{(2\pi\nu)}^2}{c^2} = \displaystyle \sum_{i=1}^{3} {n_i}^2 
{\left(\frac{\pi}{L_i}\right)}^2 + \ {\kappa}^2 \, .
\end{equation}

The equations concerning the Dirichlet conditions on the conformal coordinates 
$a$ and $b$ 
\begin{eqnarray}
c_1 \, J_{l+2} (\kappa a) + c_2 \, N_{l+2} (\kappa a) &=& 0 \\
c_1 \, J_{l+2} (\kappa b) + c_2 \, N_{l+2} (\kappa b) &=& 0 
\end{eqnarray}
have non-trivial solution only if the determinant 
\begin{equation}
J_{l+2} (\kappa a) \, N_{l+2} (\kappa b) - 
J_{l+2} (\kappa b) \, N_{l+2} (\kappa a) = 0 
\end{equation}
vanishes, thus providing the values for $\kappa$. They are hard to calculate, 
though. So we abdicate the formal treatment in favor of a more useful approach.

The graphs of these cylindrical $ J_m (x) $ and $ N_m (x) $ functions look roughly 
like oscillating sine or cosine decaying functions, with their roots asymptotically 
(large $x$) periodic. In fact, for sufficiently high argument values 
\begin{eqnarray}\label{assimptoticbehavior}
J_{l+2} (\kappa z) &\approx& \sqrt{\frac{2}{\pi \kappa z}} \cos \left( \kappa z - 
\frac{\pi}{4} (2l+5) \right) \, , \\
N_{l+2} (\kappa z) &\approx& \sqrt{\frac{2}{\pi \kappa z}} \sin \left( \kappa z - 
\frac{\pi}{4} (2l+5) \right) \, ,
\end{eqnarray}
so we can reason that 
\begin{equation}
\kappa \, ({b-a}) \approx \eta \, \pi \ \ , \ \ \eta=0,1,2,... 
\end{equation}

Then for each value of $\eta$ there are $\delta \approx {2 \, a \eta}/{(b-a)}$ 
possible values for $l$, which grants $D_{\eta}$ degeneracies due to the possible 
$l_j$ values, 
\begin{equation}\label{degenerN}
D_{\eta} = \displaystyle \sum_{l=0}^{\delta} {d_l} = 
\frac{1}{5!} (\delta+1)(\delta+2)(\delta+3)(\delta+4)(2\delta+5) \, .
\end{equation}

Analogously, since the ${\cal M}$ spatial and the ${\rm AdS}_5$ conformal 
contributions are not necessarily of the same order of magnitude, one has
\begin{equation}\label{etasep}
R (T) = \frac{c}{2V} \left[ \sum_{\eta=0} + \sum_{\eta \neq 0}
            \right] \frac{h \nu_{i\eta}}{e^{h\nu_{i\eta}/{kT}}-1} \ .
\end{equation}

Working out eq.(\ref{etasep}) for $\eta=0$ leads to the known blackbody relations, 
while the $\eta \neq 0$ contribution gives 
\begin{equation}\label{R10def}
R_{\eta}(T) \doteq \frac{c}{2V} \sum_{\eta \neq 0} \frac{ D_{\eta} \ h\nu}
{e^{h\nu/{kT}}-1} 
\end{equation}

Note that the $D_{\eta}$ degeneracies given by eq.(\ref{degenerN}) are related to 
the frequencies through eq.(\ref{spec10}). Considering the mean values 
of this dispersion relation we get
\begin{equation}
\langle {\kappa}^2 \rangle = \frac{1}{4} {\left( \frac{2\pi\nu}{c} \right)}^2 \, ,
\end{equation}
which allows the substitution $D_{\eta} \rightarrow D_{\eta}(\nu)$ with 
$\kappa = {\pi \nu}/{c}$. Then, taking the sum by an integral and considering just 
the $D_{\eta}(\nu)$ leading order contribution one gets
\begin{equation}\label{R10int}
R_{\eta}(T) = \int_{0}^{\infty} R_{\eta}(T,\nu) \, d\nu 
\end{equation}
with the corresponding spectral radiancy given by
\begin{equation}\label{R10nu}
R_{\eta}(T,\nu) = \frac{2}{5!} {\left(\frac{2 a \nu}{c} \right)}^5 \,
                      \frac{{\Omega}_{(4)} \, h c}{2 \, c^4} (b-a)
                     \frac{{\nu}^{4}}{e^{h\nu/kT}-1} \ .
\end{equation} 

Through the variable change $y=h\nu/kT$ one gets
\begin{equation}
R_{\eta}(T) = \frac{{\Omega}_{(4)}}{2 \cdot 5!} \frac{hc \,{(2a)}^6}{c^9} \left( 
\frac{b}{a} - 1 \right) {\left( \frac{kT}{h} \right)}^{10} \int_0^{\infty} 
\frac{y^9}{e^y-1} \, dy \ ,
\end{equation}
and the radiancy contribution due to these modes is 
\begin{equation}\label{R10}
R_{\eta} (T) = \sigma_{\eta} \, T^{10} \ ,
\end{equation}
\begin{equation}\label{sigma10}
\sigma_{\eta} \, = \, \frac{{\Omega}_{(4)} k c}{2 \cdot 5!} \, 
         {\left( \frac{k}{hc} \right)}^{9} 
         \Gamma(10) \, \zeta(10) \, {(2a)}^6 \left(\frac{b}{a}-1\right) \, .
\end{equation}

\vspace{1mm}
Thus, the total blackbody energy rate per unit area is 
\begin{equation}\label{RTotal10}
R(T) = \sigma_{_B} \, T^4 + \sigma_{\eta} \, T^{10} \ ,
\end{equation}
which is understood as the generalized Stefan-Boltzmann law for the 
$ \textrm{AdS}_5 \times {\cal S}^5 $ spacetime.

As before, for low temperatures the above generalization reduces to the well-known 
Stefan-Boltzmann law, while for sufficiently high temperatures the 
generalized Stefan-Boltzmann law takes on its higher-dimensional character 
$ R(T)=\sigma_{\eta} \, T^{10} $, in agreement with 
\cite{Alnes:2005ed,Cardoso:2005cd}. 

\vspace{0.1cm}
The generalized Wien's law for the full $ \textrm{AdS}_5 \times {\cal S}^5 $ is 
deduced the same way as before. For a given temperature $T$ there is a 
$\lambda_{_m}$ for which $R(T,\lambda_{_m})$ is maximum, namely 
\begin{equation}\label{gwien10}
1 - e^{-{hc/kT\lambda_{_m}}} = \frac{hc}{k T \lambda_{_m}} \, 
                              \frac{1+\epsilon_{\eta}(\lambda_{_m})}
                              {5+11\,\epsilon_{\eta}(\lambda_{_m})} \ \, ,
\end{equation}
where $\epsilon_{\eta}(\lambda) = {R_{\eta}(T,\lambda)}/{R_4(T,\lambda)}$ 
is the respective {\it wavelength radiancy relative deviation} 
\begin{equation}\label{wavelengthdeviation10}
\epsilon_{\eta} (\lambda) = \frac{\pi}{2 \cdot 5!} \left( \frac{b}{a} - 1 \right)
                                      {\left(\frac{2 \, a}{\lambda}\right)}^6 \ .
\end{equation}

Then, for large wavelengths one gets out of (\ref{gwien10}) the usual Wien's 
displacement law, while for small wavelengths it reduces to its higher-dimensional 
behavior.
\begin{equation}\label{lowgwien10}
 1 - e^{-x_{_{10}}} = \frac{x_{_{10}}}{11} \quad \;; 
\quad x_{_{10}} \equiv {\frac{h c}{k T \lambda_m}} \ .
\end{equation}

\section{Bounds on the conformal coordinate values and Graphs}

According to precise measurements \cite{codata2010,Sapritsky1995,Yoon2000,
Friedrich2000,Yoon2003}, the {\it wavelength radiancy relative deviation}, 
$\epsilon_{l,\eta}(\lambda)$, shall be no greater than $1\% \sim 2\%$ for 
${\lambda}^{\ell} = 250\,$nm at $T \approx 3000\,$K. Which 
sets the upper bound on the conformal coordinate of the $ {\cal M} \times 
{\cal S}^5 $ scenario at $z_{_M} \approx 50\,$nm, with similar limits for 
$a$ and $b$ on the $ \textrm{AdS}_5 \times {\cal S}^5 $ spacetime.

\begin{figure}[!htb]
\begin{center}

\input{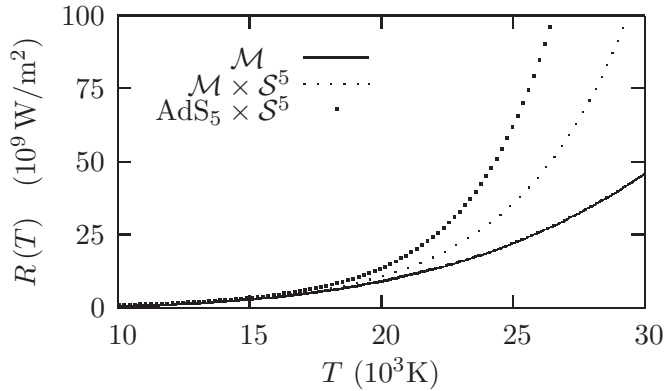}
\end{center}
\caption{Comparison between the generalized and conventional Stefan-Boltzmann laws 
for a maximum deviation $ \epsilon_{l,\eta}({\lambda}^{\ell}) $ of $1\% $.}
\label{sb}
\end{figure}

The obtained Stefan-Boltzmann law generalizations (\ref{RTotal9}) and 
(\ref{RTotal10}) are plotted in fig.(\ref{sb}) and compared with the standard 
prediction. For the considered bounds on $z$, $a$ and $b$ the discrepancy 
between these functions is just noticeable for $T>10^4$K.

\begin{figure}[!htb]
\begin{center}
\input{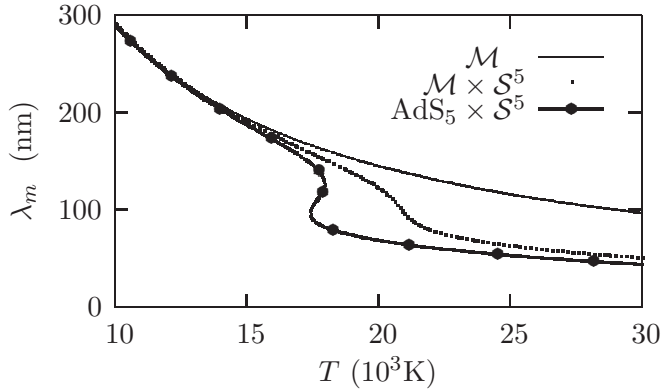}
\end{center}
\caption{Comparison between the generalized and usual Wien's displacement 
laws, for a maximum deviation $ \epsilon_{l,\eta}({\lambda}^{\ell}) = 1\% $.}
\label{wien}
\end{figure}

The generalized Wien's laws are confronted with the habitual prediction in 
fig.(\ref{wien}). The graph depicts a clear higher-dimensional signature for 
${\lambda}_m < {\lambda}^{\ell}$, where (\ref{gwien9}) and (\ref{gwien10}) 
departure from the usual behavior.

\begin{figure}[!htb]
\begin{center}
\input{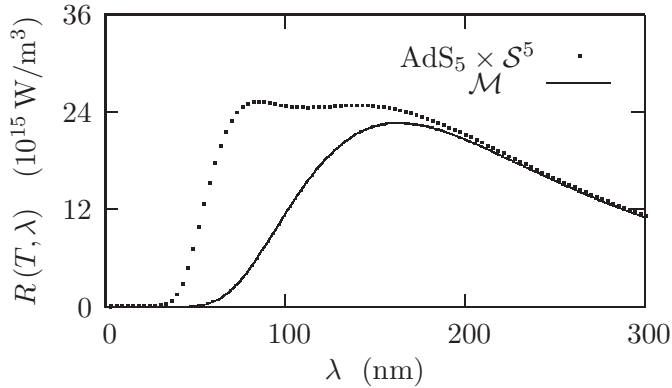}
\end{center}
\caption{$ \textrm{AdS}_5 \times {\cal S}^5 $ wavelength radiancy compared 
with the usual $ {\cal M} $ spectrum for $ \epsilon_{l,\eta}({\lambda}^{\ell}) = 
1\% $ and $T=17,75 \cdot 10^3$K.}
\label{spectral}
\end{figure}

The generalized $\textrm{AdS}_5\times{\cal S}^5$ Wien's law discloses a 
remarkable behavior on its temperature transition range. Its wavelength 
radiancy displayed in fig.(\ref{spectral}) has two distinct maxima separated 
by a minimum. This feature is accentuated in the temperature transition 
range for smaller conformal coordinate values.

\section{Conclusion and discussions}

In this work we generalize the Blackbody Radiation Laws for the compelling 
$ \textrm{AdS}_5 \times {\cal S}^5 $ spacetime. The temperature for which 
deviations in the blackbody radiation becomes relevant is inversely proportional 
to the conformal coordinate values, (\ref{RTotal9}) and (\ref{RTotal10}), while 
the wavelength for which deviations in the blackbody spectrum becomes important 
is directly proportional to them, as noted in (\ref{wavelengthdeviation9}) 
and (\ref{wavelengthdeviation10}).

The additive aspect of the generalized Stefan-Boltzmann (\ref{RTotal9}) 
(\ref{RTotal10}) and Wien's displacement (\ref{gwien9}) (\ref{gwien10}) laws 
traces back to the temperature induced split structure of (\ref{lseparation}) and 
(\ref{etasep}), the key conceptual point of our approach \cite{RamatonBoschi}. 
Thus the blackbody temperature sets up how the spacetime takes part on its 
radiative features.

Our approach yields the suitable higher-dimensional blackbody features in the 
high-energy regime and is also compatible with our empirically presumed 
$4$-dimensional spacetime, once it reproduces the observed Stefan-Boltzmann 
and Wien's displacement law in our ordinary energy scale. In the low energy scale, 
it is seen that $ \textrm{AdS}_5 \times {\cal S}^5 $ (bulk) spacetime as well as its 
Minkowskian (brane) subspace $ {\cal M} \times {\cal S}^5 $ behave effectively as an 
${\cal M}$ manifold, while in the high energy scale they behave effectively as 
${\cal M}^{1,9}$ and ${\cal M}^{1,8}$ respectively. 

Regarding the AdS/CFT correspondence, we would like to say a few words. 
It is expected a relation between normalizable $ \textrm{AdS}_5 \times {\cal S}^5 $ bulk fields 
and Yang-Mills (YM) boundary operators. Since these are vectorial bulk fields (spin 1) 
the corresponding YM operators should have the same spin. So, these operators represent spin 1 
glueballs. It is important to mention that since we imposed 
boundary conditions on the bulk fields at $z=a$ and $z=b$, with $0<a<b<\infty$, 
the original conformal symmetry of the CFT theory is broken, at least in this low energy 
regime. The high energy phase is obtained considering the limits $a\to 0$ and $b\to \infty$, in 
which the conformal symmetry is restored.

\acknowledgments

The authors are partially supported by the Brazilian agency CNPq.

\end{document}